# Shock (Blast) Mitigation by "Soft" Condensed Matter

Vitali F. Nesterenko

*Department of Mechanical and Aerospace Engineering,
University of California, San Diego,
La Jolla, CA 92093, U.S.A.*

## ABSTRACT

It is a common point that "soft" condensed matter (like granular materials or foams) can reduce damage caused by impact or explosion. It is attributed to their ability to absorb significant energy. This is certainly the case for quasistatic type of deformation at low velocity of impact widely used for packing of fragile devices. At the same time a mitigation of blast phenomena must take into account shock wave properties of "soft" matter which very often exhibit highly nonlinear, highly heterogeneous and dissipative behavior. This paper considers applications of "soft" condensed matter for blast mitigation using simplified approach, presents analysis of some anomalous effects and suggestions for future research in this exciting area.

## INTRODUCTION

"Soft" condensed matter (granular, porous materials, foams) can be successfully used for mitigation of shock wave caused by contact explosion and for reducing effects of explosion in an air. Relatively inexpensive granular materials such as iron shot, easily available as a waste from metallurgical plants, proved to be very good shock damping medium. The unusual feature of granular materials is a negligible linear range of the interaction force between neighboring particles (emphasized by term "sonic vacuum" [1]) and highly heterogeneous state under loading (deviations from average values of parameters are comparable with the averaged values). Examples of successful blast confinement using "soft" condensed matter can be found in [1]. A fact of energy absorption by "soft" matter may not preclude anomalous behavior − a significant enhancement of shock amplitude or absorbed momentum can happen instead of expected mitigation. That is why understanding of shock wave decay in granular assemblies, foams and in composite high gradient barriers is essential for the development of strong shock absorbers for violent dynamic loading from such threats like contact and air explosion.

## REACTION OF STRUCTURES ON IMPULSE LOADING

It is remarkable that global behavior of structures under impulse loading $P(t)$ (pressure in reflected shock wave) can be illustrated by a simple oscillatory system represented in Figure 1 [2]. The main parameters of this model include mass $M$ of vibrating body and massless spring with stiffness $k$. Dashpot can be added to account for dissipation. Natural period of oscillations $T = 2\pi (M/k)^{1/2} = 2\pi/\omega_n$ plays the main role in determining a type of response.

For example such model was used to design explosive chambers for localization of blast and can be used to analyze blast loading of a building frame [2-5]. In the case of thin wall spherical explosive chamber $T$ depends on the radius of chamber $R_0$, density of the material $\rho$ and its elastic properties (Young's modulus $E$ and Poisson ratio $\nu$) [3]

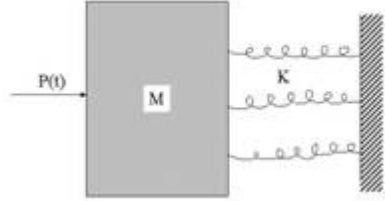

**Figure 1**. Simple model for analysis of dynamic behavior of structures under impulse loading.

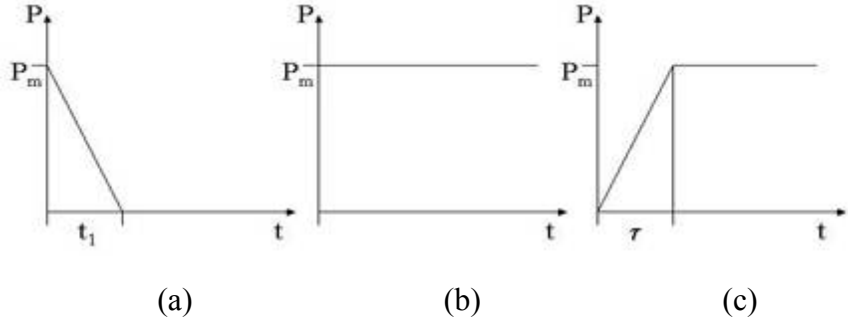

**Figure 2**. Three types of impulse loading $P(t)$: (a) triangular pulse with sharp front and duration $t_1 \ll T$, (b) step function with sharp increase of pressure and (c) ramp function with finite time $\tau$ of pressure increase. $P(t)$ represents pressure in a reflected shock wave.

$$T = 2\pi R_0 [(1-\nu)\rho/2E]^{1/2}. \tag{1}$$

More complex approaches are needed to account for observed "swinging" behavior of chamber vibrations or take into account specific geometrical features of localization devices [1].

Three types of dynamic loading relevant to our purpose are presented in Figure 2. A reaction of a model system (Fig. 1) to these loading conditions can be found using for example the convolution integral [5]. The maximum responses for each case are given below:

$$x_m = \frac{P_m \omega_n t_1}{2k} = \frac{I\omega_n}{k}, \tag{2a}$$

$$x_m = \frac{2P_m}{k}, \tag{2b}$$

$$x_m = \frac{P_m}{k}\left\{1 + \frac{1}{\omega_n \tau}\sqrt{2(1-\cos\omega_n\tau)}\right\}. \tag{2c}$$

Under the loading by a short triangular pulse (Fig. 2a) impulse of shock pressure $I$ determines the maximum displacement $x_m$ and stresses in a structure (Eq. 2a), shape of $P(t)$ has no effect at constant $I$. It means that application of "soft" condensed matter to tailor impulse shape has no sense in this limit. It is true for example in many cases for blast loading of building structures.

In case of step function (Fig. 2b), a maximum amplitude is determined by a pressure maximum $P_m$ and it is equal to two static displacements (Eq. 2b). In this case reducing amplitude of pressure $P_m$ is very desirable for purpose of mitigation. Important feature of forced vibrations excited by a step function is a shifting of their center from equilibrium position [5]. Despite that impulse of pressure can be very large, value of $I$ is not relevant in this limit.

Response to ramp function (Fig. 2c) is represented by Eq. 2c. It has a tendency to decrease with increase of ratio $\tau/T$ and in the limit $\tau \gg T$ ($\omega_n \tau \gg 1$) system responds statically ($x_m = P_m/k$), again with no influence of shock pressure impulse $I$. Reduction in maximum pressure is beneficial for blast mitigation. Equation 2b is a partial case of Eq. 2c at $\omega_n \tau \ll 1$.

We may conclude that tailoring of pressure pulse with "soft" condensed matter is useful for mitigation of blast effects if it is possible to make duration of impulse longer (actually it may be comparable) than natural period of vibration $T$. In this limit ramping of shock front may additionally reduce maximal displacement in two times in comparison with step function. From this simple analysis we observe that the desirable output of application of "soft" matter should be the transformation of impulsive type of $P(t)$ (Fig. 2a) to a long duration loading and ramping of shock front with low maximal pressure (Fig. 2c). This is possible to accomplish for explosive chambers [1] with typical diameters of few meters but practically very difficult to achieve using cladding by "soft" materials for building structures. This is because such protection did not decrease absorbed momentum and typical $T$ for global vibrations is very large for building structures in comparison with duration of shock loading pulse $P(t)$ even if "soft" matter is used.

**EXAMPLES OF BLAST MITIGATION USING "SOFT" CONDENSED MATTER**

There are two types of blast loading (Figures 3 and 4) which are dramatically different according to a duration and pressure level. Figure 3 corresponds to a contact explosion with typical level of shock pressures $P_2$ of 10 000 – 50 000 MPa depending on type of explosive and material of the wall (or in a protective barrier) and typical duration about few tens of microseconds at explosive thickness about 10 cm. Figure 4 represents air explosion where level of shock pressure and duration depend on type of explosive, its mass and distance to the wall. Typical level of pressure in reflected shock wave in this case is about 8 MPa and duration of 2 milliseconds for mass of explosive 1 kg (in TNT equivalent) detonated at distance 1 m from a wall [2]. $P_2$ and $P_3$ are shock pressures in a wall in corresponding cases. Their amplitudes can be very important for local response of structure if for example spall fracture in a wall is probable. Oscillatory global behavior of a structure may be determined by a momentum $I$ or by pressures $P_2$ and $P_3$, as was discussed earlier. The examples of successful application of "soft" condensed matter for blast mitigations in both conditions can be found in [1] with corresponding references to original papers. The most important case corresponding to the protection against damage due to a contact explosion is related to granular beds made from iron shot which is used as support table in explosive chambers. Despite of the lack of understanding of nonlinear behavior of real 3-D granular materials, they are effectively used in practice. Granular materials as cast iron shots are also used as a support for basic plate in explosive welding process.

Granular material of the same type can effectively prevent spall in reinforced concrete constructions at contact explosive loading. The layer of cast iron shot with thickness only 20 mm (particle size 3-5 mm) placed between explosive charge and a steel plate with thickness of 300 mm can prevent the fracture ("perpendicular spall") developing from surface defect [1].

The examples of stress reduction in stationary explosive chambers and in disposable structures using sawdust as a barrier are presented in Figs. 5 and 6. The observed mitigation effect is due to the change of structure loading from impulsive regime (Fig. 2a, Eq. 2a), where momentum $I$ dominates response, to a quasistatic regime under ramp function loading where pressure with a small amplitude is a main parameter. This mechanism of blast mitigation is supported by the fact that a vibrations are characterized by the shifting of their center from the equilibrium position in case where sawdust barrier was used (compare Figs. 5a and 5b).

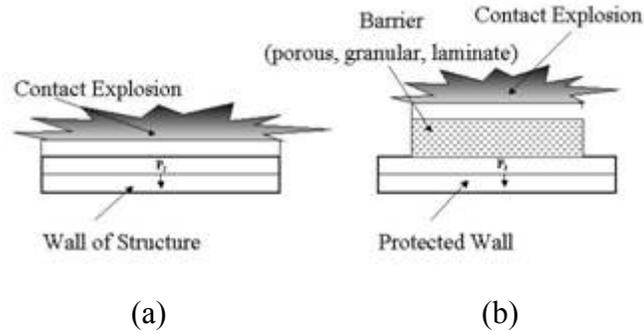

**Figure 3**. (a) Blast loading due to contact explosion without protected layer and (b) with protected layer, $P_2$ and $P_3$ are shock pressures in the wall.

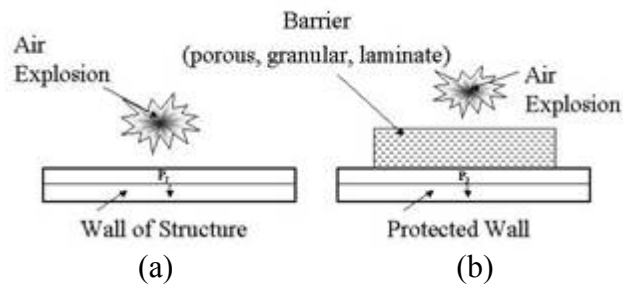

**Figure 4**. (a) Blast loading as a result of air explosion without protected layer and (b) with protected layer, $P_2$ and $P_3$ are shock pressures in the wall.

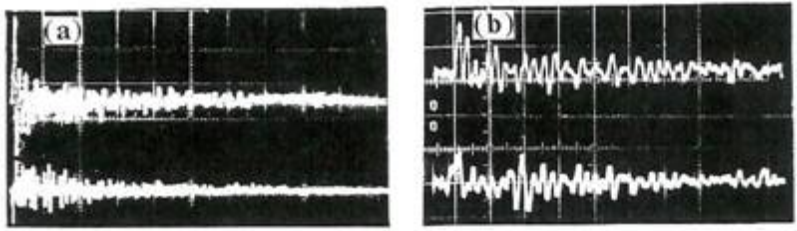

**Figure 5**. (a) Strains in the cylindrical chamber for explosion in air (RDX, 30 g) with symmetrical vibration about equilibrium position and (b) in sawdust (RDX, 60 g) with shifted center of vibrations [1]. Vertical scales 50 MPA/div, horizontal 2 ms/div (a) and 0.5 ms/div (b).

Mitigation of blast damage using disposable, light-weight structures with sawdust as a barrier is illustrated in Fig. 6. The thickness of porous barrier must be greater than about six to eight characteristic diameters of the explosive charge. Otherwise opposite effect of load amplification is possible. Using optimal porosity of porous barrier is also very important [1].

Results for polymer foams and for composite foam/solid laminar barriers can be found in [1].

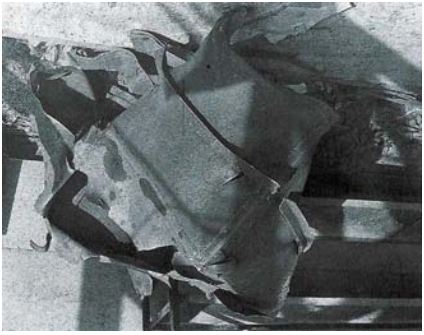
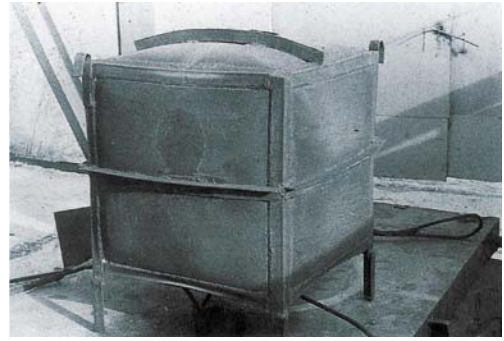

(a)             (b)

**Figure 6**. (a) Catastrophical failure of steel structure after inside explosion in air and (b) confined explosion in the same structure filled with sawdust. Explosive (RDX) mass was 0.5 kg in both cases, wall thickness of containers 3 mm, side 0.7 m, sawdust density 100 kg/m$^3$ [1].

**PROPERTIES OF "SOFT" MATTER RELEVANT TO SHOCK MITIGATION**

It is very important to establish criteria for materials properties and geometry which will ensure mitigation for specific conditions of blast loading. The fact that such materials are efficient energy absorbers does not automatically guarantee their mitigation performance. For example application of a porous layer with small thickness equal to an effective radius of explosive charge did not result in damping of strains in explosive chamber [1].

Another interesting fact is that a copper powder with high thermal conductivity and sawdust with low thermal conductivity provide the same effect according to the reduction of the wall strains despite a difference in a density of these materials more than one order of magnitude (at the same ratios of a radius of porous shell and chamber to the effective radius of explosive charge). This also means that heat conductivity of porous media does not influence the reduction of the chamber's strains. Air gap between explosive charge and porous media did not change mitigation effect. It demonstrates that heat losses due to the contact of cold powder and hot detonation products are of secondary importance, which is consistent with other observation [6].

The damping effect is mainly connected with the qualitative change of chamber loading regime caused by porous media [1]. The main parameters determining the effectiveness of blast mitigation are density, porosity and relative geometrical size of "soft" matter.

Significant improvement of mitigation capability of porous barrier can be achieved by appropriate "organization" of porous space. For example in case when structure is loaded by step function and reacts on the maximum pressure the porous barrier which ensures transition to ramped loading (Fig. 2c) can decrease the maximum stress in two times and also prevent spall in a wall of protected structure. This may be achieved using multiporous materials ( Fig. 7). This structure introduces a new scale – radius $a_0$ of large pores. The width $\Delta$ of a strong stationary shock wave in this material can be evaluated by neglecting strength of granular matrix. In frame of Carroll-Holt model $\Delta$ is described by the following equation [1]

$$\left(\frac{\Delta}{a_0}\right)^2 = \frac{4 \cdot 2^{1/3}(\alpha_{0G}-1)\,\alpha_0^2}{(2\alpha_{0G}^3 - 3\alpha_{0G}^2 + 1)} \left[1 - \left(\frac{\alpha_{0G}-1}{\alpha_{0G}+1}\right)^{1/3}\right], \tag{3}$$

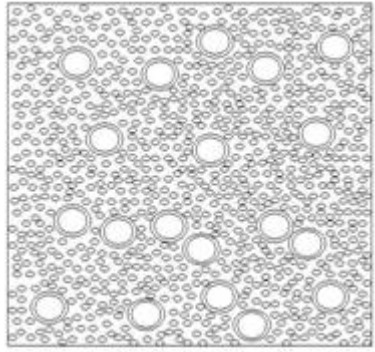

**Figure 7**. Multiporous (large cavities in granular matrix) material for shock front tailoring.

where $\alpha_{0G}$ is a porosity of multiporous material determined by ratio of initial density to the density of granular matrix (we consider a special case when large size pores are collapsed without densification of granular matrix – case of great importance for repeatable use of this material. If $\alpha_{0G}$ is about 2 the estimated $\Delta$ is about $1.2a_o$, which can be made much larger than particle size in granular matrix - the shock front width in the "normal" granular material.

Multiporous geometry can be realized in experiments by placing hollow spheres or cylinders made from low strength materials, like thin walled metal or even plastic or paper shells, in granular matrix composed, for example, from iron shot. Specific shape of pores, different from spherical or cylindrical can be preferable for initiation of collapse by organizing flow of collapsing granular materials through appropriate type of instabilities due to initial geometry of porous space. This "internal avalanche" inside multiporous granular material can be very useful for energy dissipation purposes. The dissipated energy $\varepsilon$ per unit mass in stationary shock is

$$\varepsilon = \frac{1}{2} p (V_0 - V_f), \tag{4}$$

where $V_0$ and $V_f$ are initial and final specific volumes of multiporous material and $p$ is a shock pressure. The mechanism of this dissipation can be due to large relative displacements of neighboring particles and their friction and collisions during collapse of large pores.

Another example of porosity organization can be realized in layered systems composed from granular layers of different shape (or layers of multiporous materials, Fig. 7) separated by air gaps. This may more efficiently use "internal avalanche" process for energy dissipation.

**ORDERED "SOFT" MATERIALS**

Periodically arranged systems, for example laminates, like Steel-Porolon system of metal plates and polymer foam [1], were investigated for shock mitigation. The idea was to employ a dramatic difference in acoustic impedances of steel and foam which is expected to result in quick decay of leading shock due to its multiple interactions with interfaces. Then decreasing thickness of the layers at the same overall effective density and size of system should result in more effective shock decay. But nonlinearity of material behavior results in the opposite effect for strong shocks [1, 7]. The specific nature of materials in laminate does not matter, mainly nonlinearity of their behavior is responsible for this phenomena. The mechanism of observed effect is connected with the overtaking of the leading wave by secondary shocks [1].

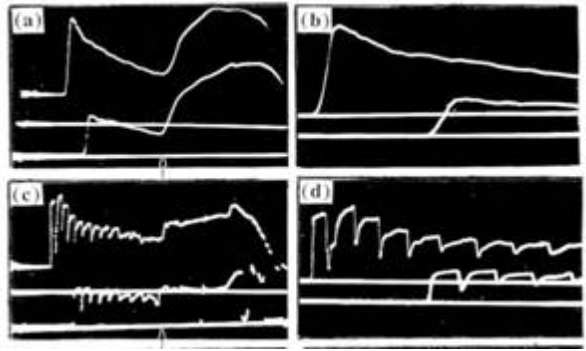

**Figure 8**. (a), (b) Particle velocity profiles in random system (glass powder) and (c), (d) in laminar ordered system at the same loading conditions: (a), (b), $L^1 = 0$ mm, $L^2 = 15$ mm; (c), (d), $\delta = 2.9$ mm, $L^1 = 1.45$ mm, $L^2 = 16$ mm. Each pair of pictures represents the same experiment. For (a), (c), - whole horizontal line equals to 100 μs; for (b), (d), - 30 μs.

The mesostruture of laminar material is important in case of laminar Steel-Porolon system. It was shown that single-cell material is more effective in comparison with the two-cell material and even with three-cell structure. The reason for this behavior is connected with nonlinear effects and secondary wave reflections. It is important that massive steel plate must face the shock to reduce energy absorption from air shock [1].

Interesting feature of ordered system is an oscillating shock wave structure causing higher amplitude on front. For different systems this may be due to a completely different mechanisms. For example, oscillating particle velocity profiles are characteristic for laminated system of glass plate and air and monotonous profiles for random system (powder) composed from glass particle with the same overall porosity at different depths $L$ (Fig. 8, [1]). The mechanism for oscillation behavior is due to a periodic origination of shock waves caused by an impact of a first moving plate on a plate in front of it followed by rarefaction wave when a leading shock inside a plate approaches a corresponding free surface. Comparison of shock profiles in two systems clearly demonstrates that "organization" of porosity can be crucial for tailoring of shock response.

Faster achievement of final equilibrium state in disordered macroscopically identical material should be mentioned (compare Figs. 8 (a), (b) and Figs. 8 (c), (d)). This is due to a qualitatively different mechanisms of dissipation in these two cases – shock/rarefaction sequences in ordered case and plastic deformation, fracture and friction in disordered system.

Another example of "soft" periodical matter with unique properties is a chain of elastic particles interacting by Hertz law which represents a case of "sonic vacuum". Such strongly nonlinear phononic materials (metamaterials) may be specifically designed and assembled to utilize their unique capabilities for signal transformation and processing [1, 8, 9, 10], including shock mitigation. One of possibilities to tailor response of such metamaterials is to use composite high density particles coated by material with low elastic modulus $E$.

Impact of "metamaterial" assembled from 20 steel particles (ø 4.75 mm), by piston with velocity 1 m/s, mass equal 30 mass of particles resulted in oscillatory shock wave presented in Fig. 9(a). For a shorter duration of loading the sequence of solitary waves can be observed [1]. Oscillatory behavior in this case is due to a repeatable elastic loading and unloading of particle contacts. If geometrically identical system of lead particles is loaded in a similar way the shock profile is monotonous (Fig. 9(b)) testifying that plastic deformation of contacts may prevent

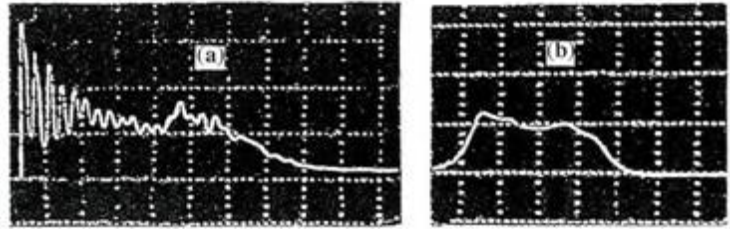

**Figure 9**. (a) Oscillatory shock profile in the system of elastic steel particles and (b) monotonous shock in the geometrically equivalent system of plastic lead particles. Vertical scales are (a) 92 N/div and (b) 18.5 N/div, horizontal scales are (a) 50 µs/div and (b) 200 µs/div.

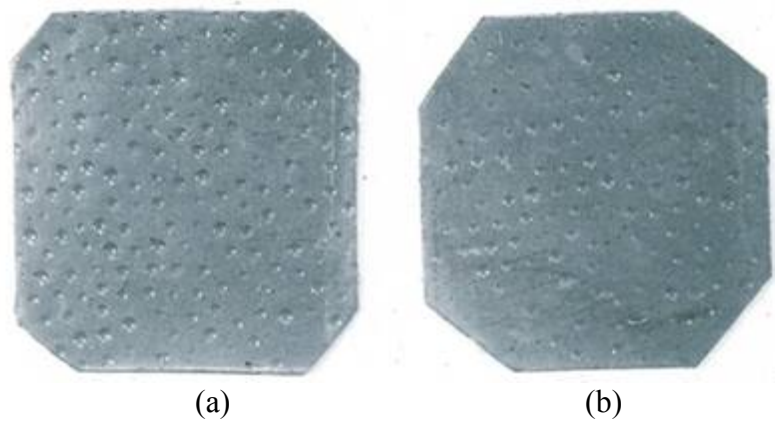

(a)     (b)

**Figure 10**. Wide range of indentation diameters demonstrates highly heterogeneous contact loading. Lead plate placed on distance 50 mm (a) and 100 mm (b) from contact with explosive.

oscillatory shock and avoid high stresses in the front.

Maximal amplitude of contact force in case of elastic particles is more than two times larger than equilibrium value. Another interesting feature is that ramp time in case of lead particles is much longer than front width of leading shock in system of elastic particles (260 and 4.5 microseconds correspondingly) and equilibrium contact force is five times smaller (difference between maximum values is more than order of magnitude!).

The random strongly nonlinear chain has properties quite different from a periodically ordered system. It does not allow the long wave analytical approach used for periodical chain. Very important differences between ordered and random system of particles were found in numerical calculations. For example, random system under piston impact has no tendency toward the uniform steady state near the piston, unlike in the case of identical sizes (Fig. 9a). The velocity amplitude at the front does not anymore represent the maximum velocity in the system like in periodic chain [1]. The important feature in this case is a decay in amplitude of pulse even in an absence of dissipative losses [1, 8, 9]. Interesting that a significant increase in the amount of particle chaotization does not lead to an enhanced damping of the velocity amplitude.

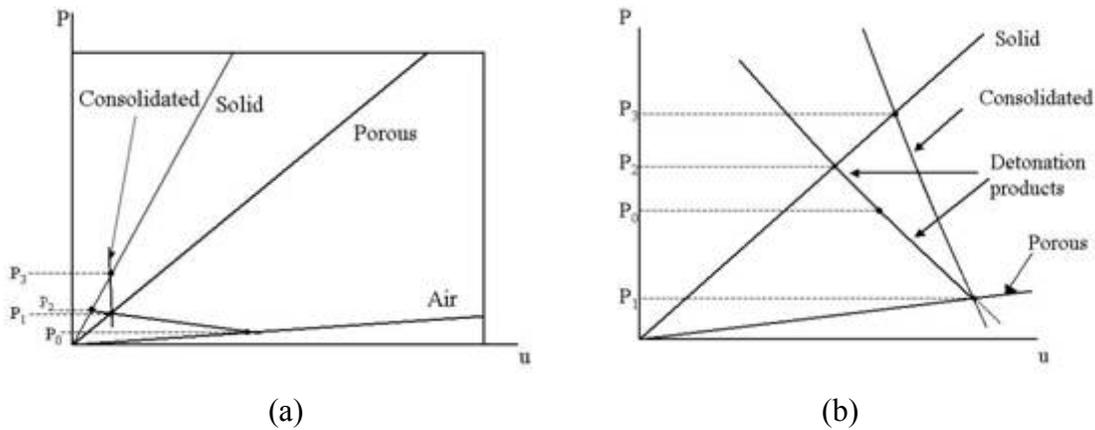

(a)                          (b)

**Figure 11**. (a) Enhancement of shock amplitude due to porous intermediate layer for air shock and (b) for contact explosion.

On the contrary, damping was smaller in the most random case. Tampered granular chains were proposed for attenuation of impulse loading [10].

Three dimensional disordered granular media attenuate shock wave in an efficient way due to a random internal mesostructure which includes one-dimensional elements − "force" chains. This highly heterogeneous structure determines material behavior even under loading by contact explosion. It is suggested by observation of broad spectrum of indentation diameters (the largest diameter is about 3 mm) of iron particles with diameter about 3 mm on adjacent lead plate (supported by massive steel plate) placed on different depths (50 and 100 mm) from surface loaded by contact explosion (Fig. 10). Cover steel plate with thickness 10 mm was placed between explosive (density $1 g/cm^3$, detonation speed 4 km/s, height 40 mm) and iron shot granular bed. Using approach similar to applied in the Brinell hardness test we find that ratio of corresponding contact forces, resulted in indentation diameters 0.5 mm and 2 mm, is about 20.

The laminar periodic system composed from heavy metal plates and light layers of foam ("springs") can be also an example of strongly nonlinear discrete system with behavior similar to the chain of elastic or plastically deformed spheres depending on properties of foam [1]. It allows tailoring of shock response by changing exponent in the power interaction law in a wider range. This exponent determines a soliton width and may qualitatively affect a shock wave profile, particularly a number of oscillations and width of leading shock wave.

**ENHANCEMENT OF SHOCK AMPLITUDE BY "SOFT" MATTER**

It is often considered that high level of energy absorption by porous matter automatically ensures effect of shock mitigation. In reality it is often observed that application of low density porous layers results in increase of shock pressure in a protected barrier [1]. The reason for this behavior under air shock loading with shock pressure in incident wave $P_0$ (Fig. 4(b)) is illustrated by self-explanatory *P*-u (shock pressure-particle velocity behind shock) diagram in Figure 11 (a).

Shock enhancement effect in case of contact explosion is illustrated by *P*-u diagram shown in Figure 11(b) ($P_0$ in this case corresponds to CJ point in explosive) and may be even more dramatic. In both cases shock pressure in a protected wall $P_3$ is significantly higher than shock pressure $P_2$ without protection layer. This enhancement can be especially severe for contact explosion with intermediate porous layer with extremely small density and small thickness.

Nature of this effect is due to the fact that shock loading of low density porous material results in a high velocity of consolidated thin layer with subsequent increase of pressure at impact of this dense material on the protected wall. Mass, linear momentum and energy conservations laws for a stationary shock ensure that high value of dissipated energy in porous media is equal to kinetic energy of densified material behind shock (if potential energy is neglected). This may cause a major difference in application of "soft" matter in quasistatic and dynamic conditions.

This shock enhancement can be avoided using attenuation of shock wave in porous material with sufficient thickness. The critical thickness $H^*$ of damping medium depends on material and on a pressure amplitude and duration of incident impulse. The equation for critical thickness of Porolon foam ensuring shock mitigation for relatively high shock pressures may be written as

$$H^* = 216 P_m^{0.6} t, \tag{5}$$

where $H^*$ is in mm, pressure in MPa and time in milliseconds [1].

## MOMENTUM AND ENERGY ENHANCEMENT BY "SOFT" MATTER

We already described a pressure enhancement which is caused by porous barriers of relatively small thickness where attenuation of shock is not developed. Even more interesting and unexpected phenomena of momentum and energy enhancement was recently observed [11] when surface of pendulum was "protected" by low density metallic foam covered by metal plate. Increase of linear momentum in case of application of foam layer with cover plate can be very substantial, up to 74% and energy increase up to 35%. This unusual result was explained [11] by evolving curvature of deformable cover plate ensuring higher value of linear momentum absorbed by pendulum from shock wave.

There is also another possibility for this behavior. It can be explained considering the simple model of the system cover plate – foam – pendulum presented in Figure 12. The natural frequency of pendulum is small enough to ensure impulse type of loading (Fig. 2(a), Eq. 2(a)). One dimensional modeling is appropriate for the first stage of motion of pendulum.

In this approach mass of foam is considered to be attached to the cover plate after impact. This is certainly a simplification, but it allows simple and quantitative illustration of the main points with reasonable quantitative agreement with experimental data. Application of conservation of linear momentum and energy to this system gives

$$I_0 = I - i,$$
$$\frac{I_o^2}{2 m_c} = \frac{i^2}{2 (m_c + m_f)} + \frac{I^2}{2 M} + E_d, \tag{6}$$

where $I_0$ is a linear momentum initially absorbed by cover plate due to air shock, $I$ is a momentum of pendulum after separation from foam and cover plate and $i$ is a final linear momentum of cover plate and foam, $m_c$, $m_f$ and $M$ are corresponding masses of cover plate, foam and pendulum, and $E_d$ is a dissipated energy due to plastic deformation of plate and foam.

It is considered that momentum absorbed by bare pendulum and by cover plate is the same and equal $I_0$. It is based on the fact that maximum velocity of cover plate (70 m/s) is about 30 times smaller than particle velocity in the incident air shock wave (amplitude about 6 MPa), so its reflection will be very much similar to a reflection from a rigid wall or from bare massive pendulum. On the first stage, the process of shock wave interaction with the system cover plate-foam-pendulum was considered as shock interaction with cover plate only. This is a reasonable simplification because interaction time of shock wave with cover plate is much shorter than time corresponding to foam deformation [11].

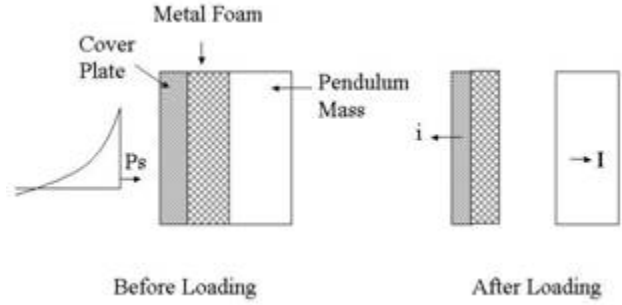

**Figure 12**. Initial and final stages of shock interaction with system cover plate-foam-pendulum.

Values of $I_0$ and $I$ are measured in experiments [11]. A value of $E_d$ can be found from Eq. 6 which is necessary to satisfy conservation laws in the proposed scenario of impulse and momentum distribution. It may be compared with evaluation of $E_d$ based on the final strain in foam and its strength. This comparison allows conclusion about the mechanisms of unexpected enhancement of blast load by foam instead of blast mitigation. From Eq. 6 energy $E_d$ is equal

$$E_d = \frac{I_0^2}{2}\left[\frac{1}{m_c} - \frac{\left(\frac{I}{I_0}-1\right)^2}{m_c + m_f} - \frac{\left(\frac{I}{I_0}\right)^2}{M}\right]. \tag{7}$$

Data corresponding to the test $D_1$ (mass of explosive charge 2.5 kg, distance from cover plate 0.5 m) are: $I_0 = 828$ Ns, $I = 1150$ Ns, $M = 935$ kg, $m_c = 11.8$ kg, and $m_f = 4.3$ kg [11]. From these data obtain value of $E_d = 25.1$ kJ using Eq. 7. Estimation of $E_d$ due to plastic deformation of foam based on its geometry (volume 0.028728 m$^3$, foam strength 1 MPa and residual strain 0.84 [11] gives value 24 kJ. An agreement between these two values supports the proposed scenario of shock loading and energy/momentum enhancement. Verification of proposed mechanism is straightforward, for example, observing effect of increasing mass of cover plate or conducting experiments with cover plate attached to pendulum with an air gap instead of layer of foam.

In this approach two major reasons are responsible for the enhancement of momentum and energy of pendulum. The first one is due to a significantly smaller mass of cover plate absorbing much larger energy from air shock than bare pendulum at a similar linear momentum (Eq.6). Second is due to a spring effect of foam which is most probably caused by an air trapped inside (foam does not allow its lateral flow). Of course, momentum absorbed by cover plate does not depend on its mass only until the plate velocity is much smaller than particle velocity in incident air shock. Momentum and energy enhancement was observed also without cover plate [11]. Part of consolidated foam facing shock could play a role of cover plate in this case (see Fig. 3.16 in [1] and explanation for double shock loading on the wall covered by layer of sawdust).

**BLAST MITIGATION BY PERFORATED STRUCTURES**

Cladding of building structures by any "soft" materials can not be expected to reduce significantly its global deformation due to a very low natural frequency of structure vibrations.

Mitigation of local shock effects (for example spall) due to contact explosion still can be accomplished by intermediate layers [1]. One approach to blast mitigation and to a reduction of global response of building can be connected with perforated structures placed in front of protected buildings, but not attached to them [1, 12]. They represent some type of "porous material" which may alter transmitted shock characteristics and subsequently reduce blast load on a building. The main idea to employ them is to force gas flow through repeated compression and rarefaction cycles with changing of flow directions, causing enhanced dissipation of air shock energy. Results for this type of blast mitigation barriers and original references can be found in [1]. Unexplored possibility is filling of cavities in such perforated structures by "soft" condensed matter like granular material or with water droplets. Then forced flow of granular material or air/water suspension through many cycles with changing of flow directions may cause more enhanced dissipation of shock energy in comparison with air. Melting of condensed matter and evaporation can provide additional sources for dissipation of energy of air shock.

**CONCLUSION**

"Soft" condense materials can be successfully used for shock/blast mitigation. Enhancement of loading amplitude or even energy and momentum is also possible specifically for low density materials if shock provides substantial densification of foam or crushing of low density structure. It is probable for shock loading with amplitude higher than crushing pressure of foam or some light weight structure and for protective layers with small dimensions where shock attenuation is not effective. This is due to the fact that conservation laws for stationary shock ensure value of dissipated energy equal to kinetic energy of densified material behind shock (if potential energy is neglected). The impact of this material moving with high velocity with protected structure and subsequent backward motion due to reflection from a wall may result in an enhanced value of pressure, momentum and energy absorbed by structure. Strongly nonlinear materials may be specifically designed for impulse transformation to utilize unique capabilities of materials with properties similar to "sonic vacuum". Investigation of shock wave dynamics in such assembled metamaterials is a fascinating subject for future research.

**REFERENCES**

1. V.F. Nesterenko, *Dynamics of Heterogeneous Materials*, Springer-Verlag, New York, 2001.
2. P.D. Smith and J.G. Hetherington, *Blast and Ballistic Loading of Structures*, Butterwoth-Heinemann Ltd., Oxford, 1994.
3. W.E. Baker, *J. Applied Mechanics, Transactions of the ASME*, March, 139 (1960).
4. A.F. Demchuk, *J. of Applied Mechanics and Technical Physics*, **9**, 558 (1968).
5. S.S. Rao, *Mechanical Vibrations*, 3$^{rd}$ edition, Addison-Wesley, Inc., 1995.
6. R. Raspet, P. B. Butler, and F. Jahani, *Applied Acoustics*, **22**, 243 (1987).
7. D.J. Benson, and V.F. Nesterenko, *J. Appl. Phys.* **89**, 3622 (2001).
8. V.F. Nesterenko, in *Granular State*, edited by S. Sen and M.L. Hunt (Mater. Res. Soc. Symp. Proc. **627**, Warrendale, PA, 2001) pp. BB 3.1.1 – 3.1.12.
9. M.Manciu, S. Sen, and A.J. Hurd, *Physica* D, 157, 226 (2001).
10. S.Sen, see paper in this Proceedings.
11. A.G. Hanssen, L. Enstock, and M. Langseth, *Int. J. Impact Engineering.* **27**, 593 (2002).
12. G.S. Grigoriev, V.E. Klapovskii, V.E. Koren'kov, V.N. Mineev, and E.S. Shakhirdzhanov, *Combustion, Explosion, and Shock Waves*, July, 98 (1987).